%% file: hadron2011.tex
\documentclass[a4paper,11pt]{article}

\usepackage{contribution}

\input{econfmacros}
\input{contributionmacros}

\begin{document}

\input{contribution}

\end{document}

%% file: econfmacros.tex
\newcommand{\weblink}[2][]{%
    \ifthenelse{\equal{#1}{}}%
    {\textnormal{\url{#2}}}%
    {\textnormal{\href{#2}{#1}}}%
}

\def\beq{\begin{equation}}
\def\eeq#1{\label{#1}\end{equation}}
\def\eeqn{\end{equation}}

\def\beqa{\begin{eqnarray}}
\def\eeqa#1{\label{#1}\end{eqnarray}}
\def\eeqan{\end{eqnarray}}

\let\bar=\overbar

\def\Dslash{\not{\hbox{\kern-4pt $D$}}}
\def\dslash{\not{\hbox{\kern-2pt $\del$}}}

\def\msb{{\bar{\ssstyle M \kern -1pt S}}}

%% file: contributionmacros.tex
\newcommand{\contribution}[7][]{%
  \clearpage
  \thispagestyle{plain}
  \ifthenelse{\equal{#1}{}}
  {\hypersetup{pdftitle={#2}}}
  {\hypersetup{pdftitle={#1}}}
  \hypersetup{pdfauthor={{#3} {#4}}}
  {\centering\normalfont\LARGE\bfseries\sffamily #2 \par\nobreak}
  \lhead{}
  \chead{%
    \textit{\footnotesize XIV International Conference on Hadron Spectroscopy
      (\weblink[\textit{hadron2011}]{http://www.hadron2011.de}), 13-17 June 2011, Munich, Germany}%
  }
  \rhead{}
  \bigskip
  \begin{center}
    {#3} {#4}\ifthenelse{\equal{#6}{}}{}{\footnote{\weblink[#6]{mailto:#6}}}
    \ifthenelse{\equal{#7}{}}{}{#7} \\
    \textit{#5}
  \end{center}
  \bigskip
}

\renewcommand{\abstract}[1]{%
  \begin{center}
    \begin{minipage}{0.85\textwidth}
      \begin{footnotesize}
        #1
      \end{footnotesize}
    \end{minipage}
  \end{center}
  \bigskip
}

%% file: contribution.tex
%
%
%
%
%
{  

\makeatletter
\@ifundefined{c@affiliation}%
{\newcounter{affiliation}}{}%
\makeatother
\newcommand{\affiliation}[2][]{\setcounter{affiliation}{#2}%
  \ensuremath{{^{\alph{affiliation}}}\text{#1}}}
%

\contribution[Short Title]
{ Nucleon Resonance Electrocouplings from the CLAS Data on Exclusive Meson Electroproduction off Protons}
{Victor I.}{Mokeev}  
{\affiliation[Jefferson Lab, USA]{1} \\
 \affiliation[Skobeltsyn Nuclear Physics Institute at Moscow State University, Russia]{2} \\
 \affiliation[Yerevan Physics Institute, Armenia]{3}}
 {mokeev@jlab.org}
{\!\!$^,\affiliation{1}, \affiliation{2}$, Inna G. Aznauryan\affiliation{3}, and Volker D. Burkert \affiliation{1}}

%

\abstract{%
  $\gamma_{v}NN^*$ transition helicity amplitudes (electrocouplings) of several prominent excited proton states
 are determined for the first time in independent analyses of $\pi^+n$, $\pi^0p$, and
 $\pi^+\pi^-p$  electroproduction off protons. 
 Analysis of  $\pi^+\pi^-p$ electroproduction has extended considerably information on electrocouplings of high lying N* 
 states, which decay preferentially to the $N\pi\pi$ final states.
}
%

\section{Introduction}

The studies of nucleon resonance structure from the data on different exclusive meson electroproduction channels off nucleons represent
an important direction in the $N^*$ program with the CLAS detector with the primary objective of determining electrocouplings
of all prominent excited proton states in a wide area of photon virtualities $Q^2$<5.0 GeV$^2$ \cite{Bu11}. 
In this paper we present the results on $N^*$ electrocouplings obtained in independent analysis of $\pi^0p$, $\pi^+n$, and
$\pi^+\pi^-p$ electroproduction off protons.

\section{Evaluation of $N^*$ electrocouplings from exclusive meson electroproduction data}
\label{models}
The $\pi^+n$, $\pi^0p$, and $\pi^+\pi^-p$ exclusive channels are major contributors to meson electroproduction off
protons in $N^*$ excitation region. They are sensitive to $N^*$ contributions and account for $\approx$ 90\% of
meson electroproduction cross section. Non-resonant contributions in these channels are different, while 
$N^*$ electrocouplings remain the same, since resonance electroproduction and hadronic decay amplitudes are independent. 
Therefore, consistent
values of
$N^*$ electrocouplings determined from different major meson electroproduction channels strongly
support a reliable extraction of these fundamental quantities.


\begin{figure}[htb]
  \begin{center}
  \includegraphics[height=.27\textheight]{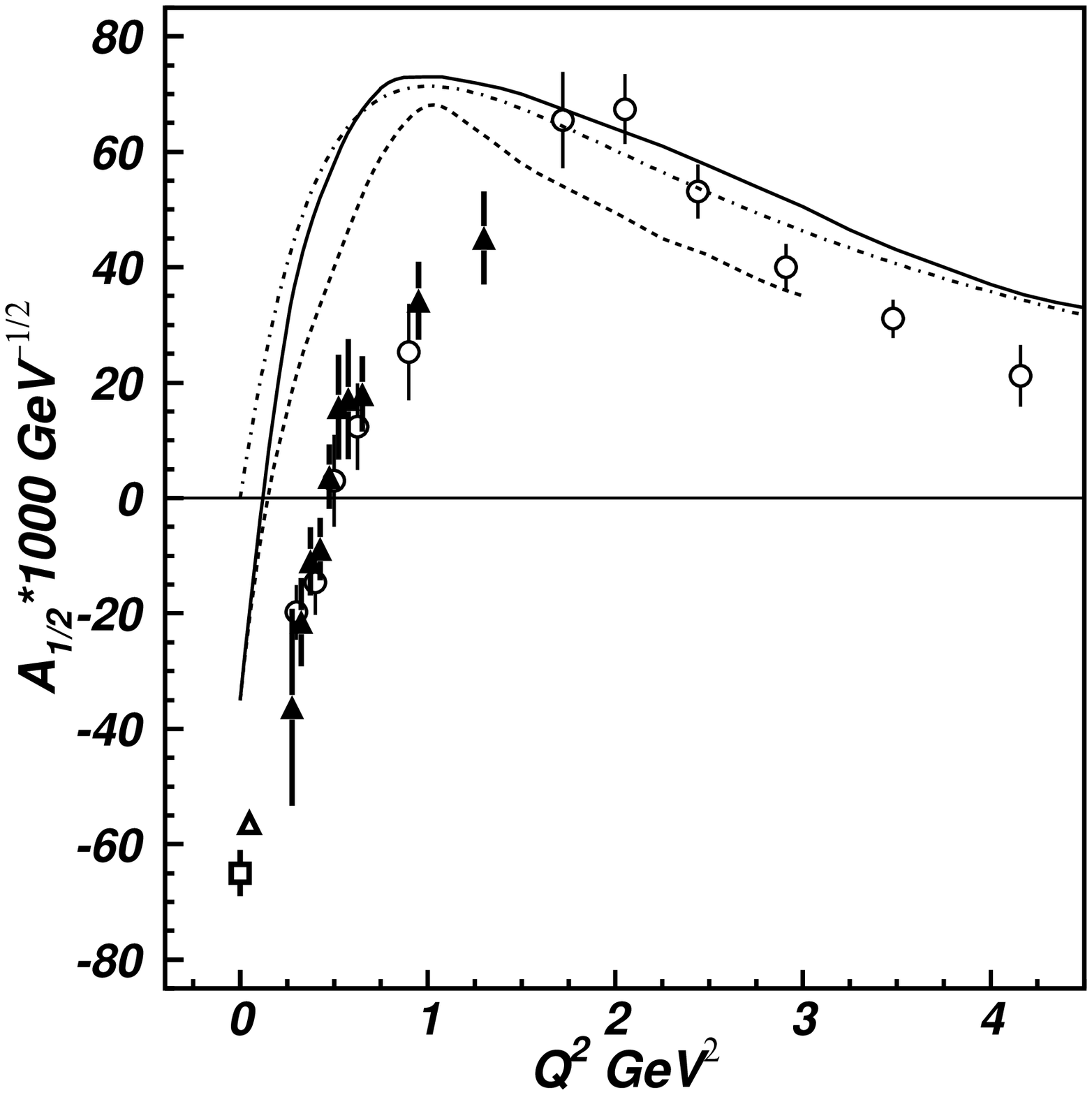}
  \includegraphics[height=.27\textheight]{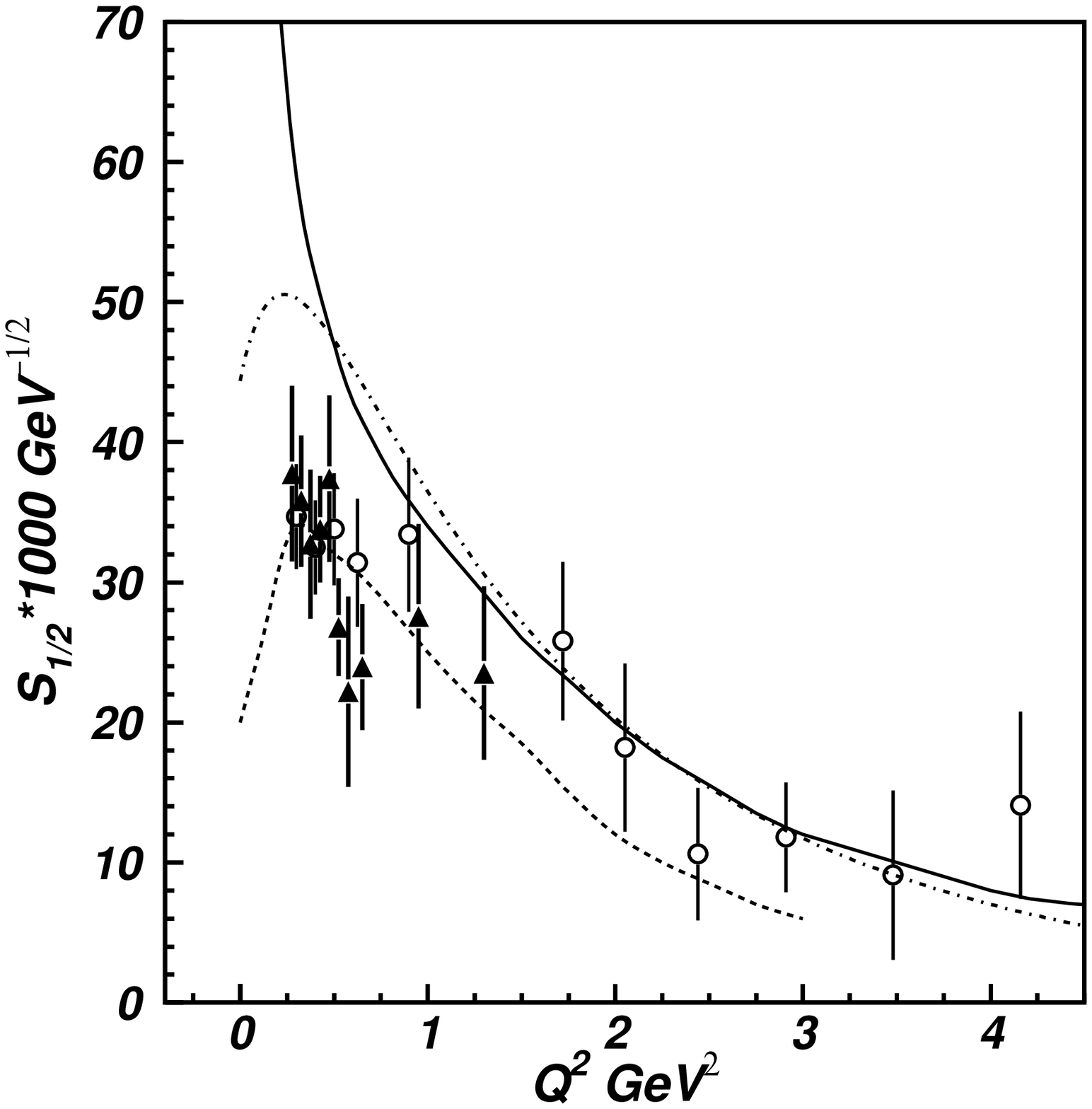}
  \caption{Electrocouplings of the $P_{11}(1440)$ resonance  determined in independent analyses of the CLAS 
  data on $N\pi$ (circles) and $\pi^+\pi^- p$ (triangles)
  electroproduction off protons. Square and triangle at $Q^2$=0 correspond to RPP \cite{rpp} and 
  the CLAS $N\pi$  \cite{Dug09} photoproduction
  results, respectively. The results of relativistic light-front quark models \cite{Az07,Ca95} are shown by solid and dashed
  lines, respectively. Results of the covariant valence quark-spectator diquark model \cite{Ra10} 
  are shown by the dashed dotted line. }
  \label{p11}
   \end{center}
\end{figure}

The CLAS data considerably extended information on $\pi^{+}n$, $\pi^0p$ electroproduction off protons. 
A total
of nearly 120000 data points on unpolarized differential cross sections, longitudinally polarized beam asymmetries, and
longitudinal target and beam-target asymmetries were obtained with almost
complete coverage of the accessible phase space \cite{Az09}. The data were analyzed within the framework of two conceptually
different approaches: a) the unitary isobar model (UIM) and b) a model, employing dispersion relations
\cite{Az03,Az05}. The two approaches provide good  description of the $N\pi$ data in the entire range 
covered by the CLAS
measurements: $W$ $<$ 1.7 GeV and $Q^2$ $<$ 5.0 GeV$^2$, resulting in $\chi^2$/d.p. $<$ 2.0.

Nine independent one-fold-differential and fully-integrated $\pi^+\pi^-p$ electroproduction 
cross sections of protons are determined from the CLAS measurements \cite{Ri03,Fe09} in 131 bins of $W$ and $Q^2$ in a mass range 
$W$ $<$ 2.0 GeV, and with
photon virtualities of 0.25 $<$ $Q^2$ $<$ 1.5 GeV$^2$.
Analysis of these data within framework 
of the meson-baryon JM reaction model \cite{Mo09,Mo11} 
allowed us to establish all essential contributing mechanisms from their manifestation in 
the measured cross sections. Reasonable data description makes it
possible to provide a reliable separation between resonant and non-resonant contributions needed for
extraction of $N^*$ electrocouplings from $\pi^+\pi^-p$ electroproduction data.

\section{Results and discussion}
Electrocouplings of the $P_{11}(1440)$, $D_{13}(1520)$ states have become available from independent analyses 
of the CLAS data on $\pi^+n$, $\pi^0p$ ($Q^2$ $<$ 5.0 GeV$^2$), and $\pi^+\pi^-p$ ($Q^2$ $<$ 1.5 GeV$^2$) 
electroproduction channels \cite{Az09,Mo11}. Their values obtained from these major meson electroproduction channels with
different non-resonant mechanisms are in a good agreement. As an example, electrocouplings of the $P_{11}(1440)$ state are shown in
Fig.~\ref{p11}. Consistent results on $N^*$ electrocouplings 
demonstrate that the reaction models \cite{Az05,Az07,Mo09} mentioned in the Section~\ref{models} provide reliable 
evaluation of these fundamental quantities. It makes possible to determine 
electrocouplings  of all resonances that decay preferentially 
to the either $N\pi$ or $N\pi\pi$ final states analyzing independently the $N\pi$ or 
$\pi^+\pi^-p$ electroproduction channels.
%


  \begin{figure}[htb]
      \begin{center}
  \includegraphics[height=.23\textheight]{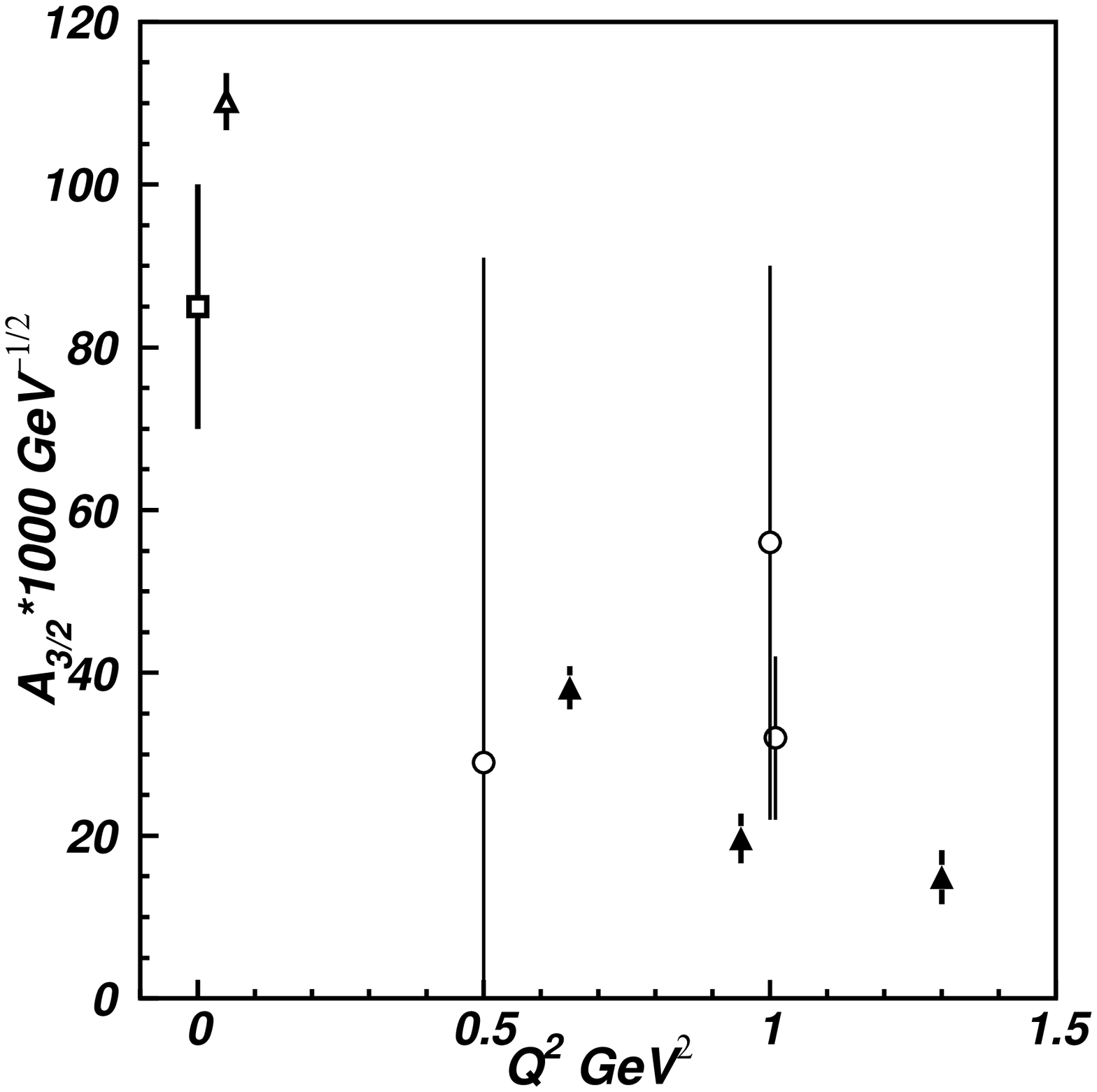}
  \includegraphics[height=.23\textheight]{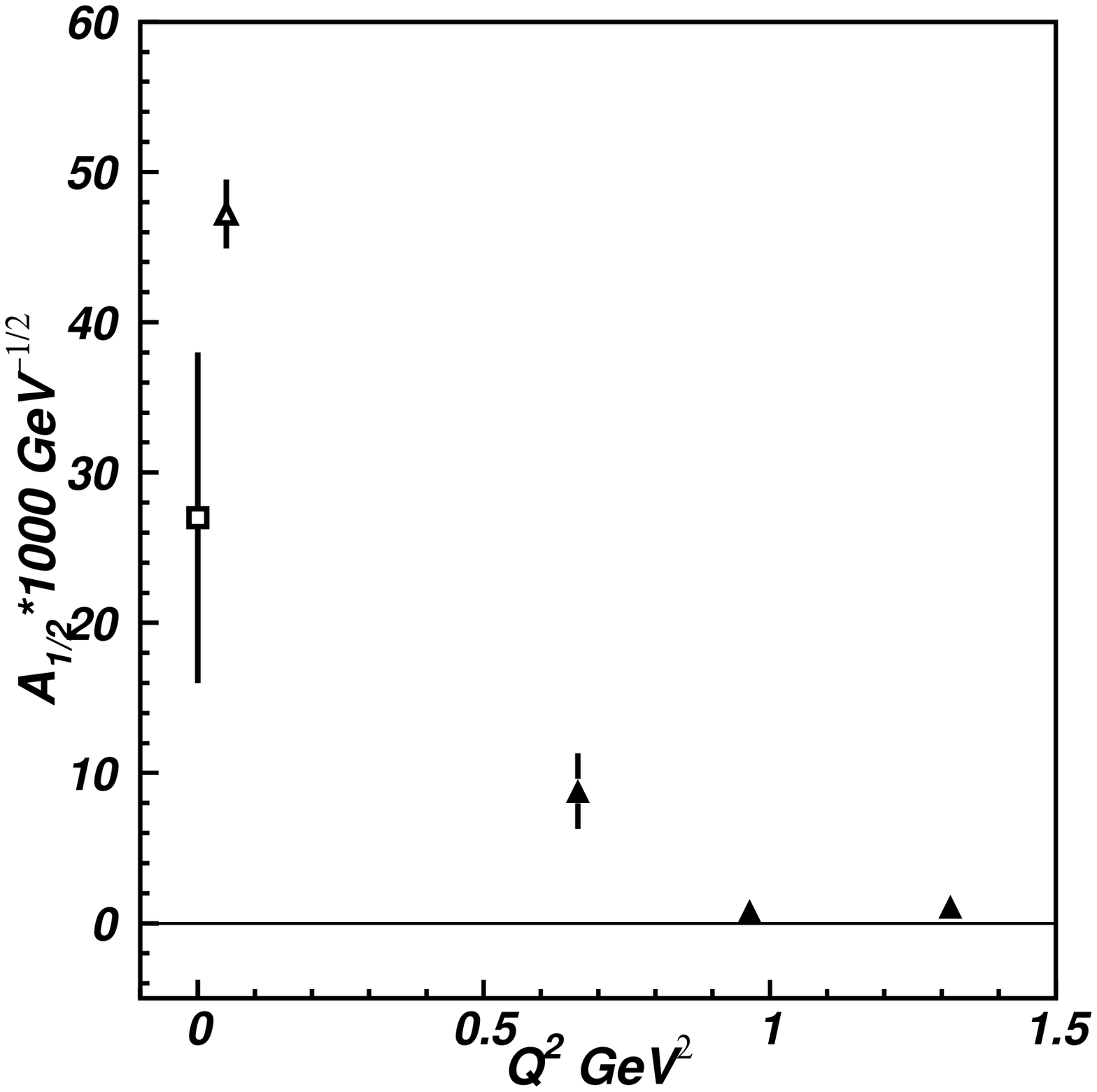}
  \includegraphics[height=.23\textheight]{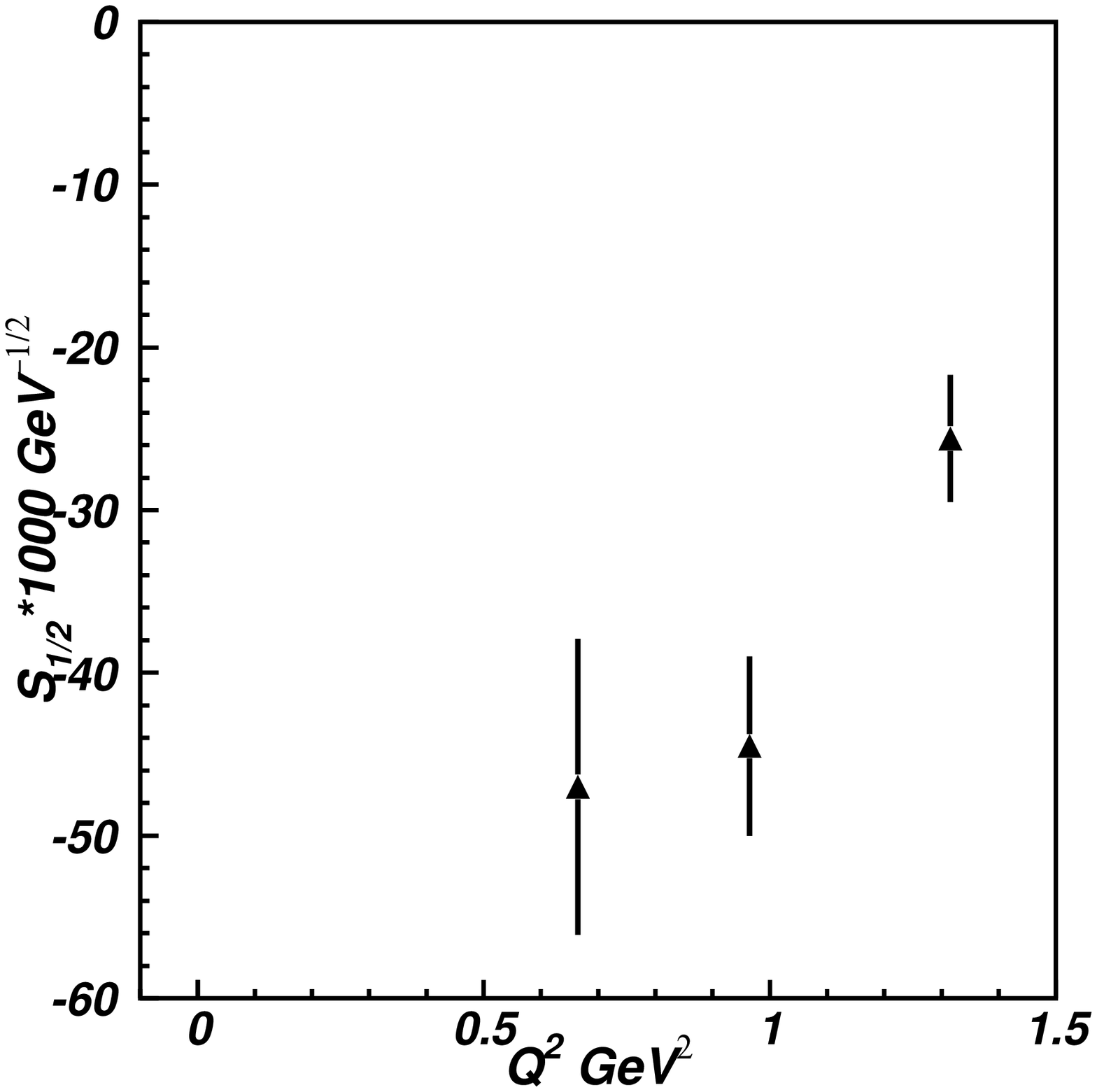}
  \caption{Electrocouplings of $D_{33}(1700)$ (left) and $S_{31}(1620)$ (middle and right)  resonances  
  from analyses of the CLAS data on $\pi^+\pi^- p$  \cite{Ri03,Fe09} and world data \cite{Bu03} on $N\pi$
  electroproduction off protons. Symbols are the same as in Fig.~\ref{p11}.}
  \label{d33s31}
    \end{center}
\end{figure}

 Preliminary results on electrocouplings of $S_{31}(1620)$, $S_{11}(1650)$, $F_{15}(1685)$, $D_{33}(1700)$ and $P_{13}(1720)$ 
 states were
 obtained from the CLAS $\pi^+\pi^-p$ electroproduction data \cite{Ri03}. The CLAS results provide accurate data on 
 the $Q^2$-evolution of the transverse 
 electrocouplings and the first information on the longitudinal
 electrocouplings of all the above mentioned excited proton states. Several examples are shown in Fig.~\ref{d33s31}. 
 A dominance of longitudinal $S_{1/2}$ electrocoupling is observed in  
 electroexcitation of $S_{31}(1620)$ state at $Q^2$ $>$ 0.5 GeV$^2$ (see Fig.~ ~\ref{d33s31}).

 The CLAS results on electrocouplings of prominent resonances stimulated the development of $N^*$ 
 structure models \cite{CRnst11,CL12}. The analysis of resonance electrocouplings within the framework of light front \cite{Az07,Ca95} and 
 quark-spectator diquark \cite{Ra10}  models, complemented by the coupled channel approach \cite{Lee08} demonstrate 
 that the structure of $N^*$ states in a mass range $W$ $<$ 1.6 GeV is determined by a combined contribution of an internal core of three dressed quarks and an external meson-baryon 
 cloud. The recent studies in the light-front quark model \cite{Bu11a} revealed an important role of dynamical mass and structure of
 dressed quarks in $Q^2$-evolution of of $N^*$ electrocouplings. Furthermore, 
 two conceptually different approaches of QCD-Dyson-Schwinger equations \cite{CR11,CRl1a}
 and  Lattice QCD \cite{Ed11,Li09,Br09}-are making progress toward the description of 
 $N^*$ electrocouplings from the first  principles of QCD.



%

}  
